\documentclass[aps,pre,preprint,groupedaddress]{revtex4}
\usepackage{graphicx}
\usepackage{booktabs}
\usepackage{topcapt}
\usepackage{epsfig}
\usepackage{bm}

\begin{document}


\title{ Liquid crystal phase and waterlike anomalies in a core-softened shoulder-dumbbells system}


\author{Alan Barros de Oliveira$^{1}$, Eduardo B. Neves$^2$, Cristina Gavazzoni$^3$,
Juliana Z. Paukowski$^3$, Paulo A. Netz$^4$, Marcia C. Barbosa$^3$}
\affiliation{ $^1$ Departamento de F\'{i}sica, Universidade Federal de Ouro Preto,
Ouro Preto, MG, 35400-000, Brazil \\
$^2$ Petrobr\'as, Av. Elias Agostinho, 665,  OP-EN-Imbetiba, Maca\'e, RJ, 27913-350,
Brazil. \\
$^3$ Instituto de F\'{\i}sica, Universidade Federal do Rio
Grande do Sul, Porto Alegre, RS, 91501-970, Brazil.\\
$^4$ Instituto de Qu\'{i}mica, Universidade Federal do Rio Grande do Sul,
Porto Alegre, RS, 91501-970, Brazil.}






\date{\today}

\begin{abstract}
Using molecular dynamics we investigate the thermodynamics, dynamics and
structure of 250  diatomic molecules interacting by a core-softened potential.
This system exhibits thermodynamics, dynamics and structural anomalies:
a maximum in density-temperature plane at constante
pressure  and maximum and minimum points in the diffusivity and translational 
order parameter against density at constant temperature.
Starting with very dense systems and decreasing density the mobility
at low temperatures first increases, reach a maximum, then decreases, reach
a minimum and finally increases. In the 
pressure-temperature phase diagram the line of maximum 
translational order parameter is located outside the line of 
diffusivity extrema that is enclosing the temperature of
maximum density line. 
We compare our results with the monomeric system showing that the 
anisotropy due to the dumbbell leads to a much larger 
solid phase and to the appearance of a liquid crystal phase. 
\end{abstract}


\maketitle
\section{Introduction}

Several anomalies present in liquid water 
are believed to be due the formation and disruption
of hydrogen bonds. Specific angles and distances
between water molecules are necessary 
to form such bonds, what leads 
to a dramatic competition between open low-density and
closed high-density structures, depending
on the thermodynamic state of the liquid.
Although in water this competition actually 
acts in the second neighbors distances\cite{Kr08}, it
could be mimetized as a competition between 
two preferred interparticle distances in a spherically
symmetrical intermolecular interaction potential.  Based on 
this mechanism, isotropic systems where particle interact
through core-softened potentials were considered
for studying the  odd behavior of water~\cite{Sc00,Fr01,
Bu02,Bu03,Sk04,Fr02,Ba04,
Ol05,He05a,He05b,He70,Ja98,Wi02,Ku04,Xu05,Ne04,Ol06a,Ol06b,Ol07,
Ol08a,Ol08b,Ol09,Fo07,Al09,Bu04a,Wi06}.
 
One of the biggest challenges for adopting these models
to describe  water and systems that exhibit the 
anomalies present in water 
is to reproduce some aspects of the anomalous behavior \cite{De96,Mi98},
such as the existence of a temperature of maximum density (TMD), 
liquid-liquid phase transition or non-monotonic diffusion
behavior with respect to isothermal pressure variation \cite{Ne01,Er01,Ne02b}.
 
The recently proposed core-softened shoulder
potential \cite{Ch96b,Ne04,Ol06a,Ol06b} reproduces the density
and diffusion anomalies. 
This potential can represent in an effective and 
orientation-averaged way the interaction between water pentamers 
characterized by the presence of
two structures -- one open and one closed -- discussed above.
Similarly, the thermodynamic and dynamic anomalies result from the
competition between the two length scales associated with
the open and closed structures. The open structure is favored
by low pressures and the closed structure is favored by
high pressures, but only becomes accessible at sufficiently high
temperature. 

Simple pair potentials are particularly interesting
because they are computationally cheaper than molecular models as well as
amenable under analytical treatments \cite{Eg08a,Eg08b,Ol06a}. For example,
the recent perturbation theory developed by Zhou \cite{Zh06,Zh08,Zh09} 
is highly promising on attacking these sort of potentials.

In spite of core-softened potentials have been mainly used
for modeling water \cite{Xu06,Ol05,Ol06a,Ol06b,Wi06,Ya05,Ya06,Fr07a,Ol09}, many other
materials present the so called  water-like anomaly behaviour.
For example, density anomaly was found experimentally in 
Se$_x$Te$_{1-x}$,\cite{Th76} and Ge$_{15}$Te$_{85}$.\cite{Ts91}  
Liquid sulphur displays a sharp minimum in the density\cite{Sa67},
related to a polymerization transition \cite{Ke83}.
Waterlike anomalies were also found in simulations for silica,\cite{An00,Ru06b,Sh02,Po97}
silicon\cite{Sa03} and BeF$_2$,\cite{An00}.

In this sense, it is  reasonable  to use core-softened
potentials as the building blocks of a broader class of materials
which we can classify as anomalous fluids. In principle
this would allow us to fabricate  anomalous liquids by
imposing interparticle core-softened potentials. For instance, 
in principle one could build a polymer that 
diffuses faster under higher pressures what would
be a quite interesting property for manufacturing plastic
materials. But polymers are anisotropic systems and 
up to now the literature
on core-softened potentials leading to water-like anomalies
is restricted to spherical symmetric systems. Would the addition
of an anisotropy in one direction to a system of particles interacting
through a core-softened potential still maintain the anomalies present
in the spherical symmetric system? Which new features the 
anisotropic  system would exhibit? 

Here we address these questions by studying the pressure-temperature
phase diagram of  a model system made
by rigid dimers. Each particle in the dimer interact with the particles
in the other dimers through a core-softened potential. The results
obtained using the dumbbell system are compared with pressure-temperature
phase-diagram of the monomeric system interacting by the same
core-softened potential employed in the dimeric case.

Obviously anisotropic systems are not only 
computationally  more complicated but also 
the addition of an extra
degree of freedom yields
a richer phase-diagram. For instance,
diatomic particles interacting through a 
 Lennard-Jones potential \cite{Kr96}
exhibit a solid phase that occupies higher pressures and temperatures
in the pressure-temperature phase diagram. In the case 
of the solid phase, the diatomic particles exhibit two 
close-packed arrangements instead of one observed in the 
monoatomic Lennard-Jones \cite{Ve03}. 
Therefore, we also expect the dimeric system interacting
through two length scales potential might have a 
phase-diagram with a larger solid phase
region in the pressure-temperature phase diagram
than the one occupied by the solid phase in
the monomeric system. Since in many cases the 
anomalies are close to the solid-liquid phase transition 
 \cite{Ol08a,Ol09} the dumbbell model studied
in this paper might have the 
anomalous region in the pressure-temperature
phase diagram uncovered by the solid phases.
We shall check if this is the case in this paper.

The remaining of this manuscript goes as follows. 
In sec. \ref{model} the model is introduced and the
simulation details are presented. In sec. \ref{results} the 
results are shown and conclusions are found in sec. \ref{conclusion}.

\section{The model} \label{model}

\begin{figure}
\includegraphics[clip=true,scale=0.3]{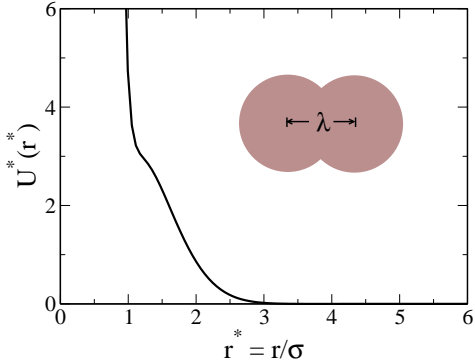}
\caption{Effective potential versus distance in reduced units.}
\label{cap:shdumb}
\end{figure}

The model consists of $N/2$ dimeric
molecules (dumbbells) formed each by two spherical particles of diameter  $\sigma,$
linked rigidly
in pairs with the distance $\lambda$ between their
centers of mass as depicted in figure \ref{cap:shdumb}.
Each particle within a dimer interacts with all
particles belonging to other dimers with the
intermolecular continuous shoulder potential \cite{Ol06a} given by
\begin{equation}
U^{*}(r)=4\left[\left(\frac{\sigma}{r}\right)^{12}-
\left(\frac{\sigma}{r}\right)^{6}\right]+
a\exp\left[-\frac{1}{c^{2}}\left(\frac{r-r_{0}}{\sigma}
\right)^{2}\right]\; .
\label{eq:potential1}
\end{equation}

This potential can represent a whole family of
two length scales  intermolecular interactions, from a deep double
wells potential \cite{Ch96b,Ne04} to a 
repulsive shoulder \cite{Ja98}, depending on the choice of the values
of $a$, $r_{0}$ and $c$. 

This potential for the monomeric system was studied
with a very small attractive region,  with
$a=5,$ $r_{0}/\sigma=0.7$ and $c=1$ so the liquid-gas unstable
and metastable region would be avoided \cite{Ol06a,Ol06b}. 
This potential  has  two length scales within 
a repulsive ramp followed by a very small attractive well.

Here we  explore the same interaction potential analyzed
for the monomeric system, but for dumbbells. In particular
we will assume $\lambda/\sigma=0.2$.
In order to study the equilibrium pressure-temperature 
phase diagram, we use molecular dynamics simulations to obtain the
pressure as a function of 
temperature along isochores,  diffusion constant as
a function of density and temperature and the behavior
of the structure as a function of temperature and pressure.

We performed molecular dynamics simulations in the canonical ensemble
using $N=500$ particles (250 dimers) 
in a cubic box of volume $V$ with periodic boundary conditions
in the three directions,
interacting with the intermolecular potential described above.
The number density of the system is then $\rho=N/V$.
The cutoff radius was set to $5.5\sigma$.
Pressure, temperature, density, and diffusion are calculated in
dimensionless units:
\begin{eqnarray}
T^{*}&\equiv & \frac{k_{B}T}{\epsilon}\; \nonumber \\
\rho^{*}&\equiv &\rho \sigma^{3} \nonumber \\
P^*&\equiv& \frac{P \sigma^{3}}{\epsilon} \nonumber \\
D^*&\equiv& \frac{D(m/\epsilon)^{1/2}}{\sigma}\; .
\end{eqnarray}

In some state points we also carried out simulations
with the same model but  with 1000 (500 dimers) 
and 2000 (1000 dimers) particles using the Large-scale Atomic/Molecular Massively Parallel Simulator (LAMMPS)  \cite{lammps}
with essentially the same results.
Further simulation details are discussed elsewhere \cite{Ol06a}.

Preliminary simulations
showed that depending on the chosen temperature and density
the system was in a fluid phase but became metastable
with respect  to the solid phase. In order to locate
the phase boundary between the 
solid and the fluid phases two sets of simulations were carried out, one
beginning with the molecules in a ordered crystal structure and
the other beginning with the molecules in a random, liquid,
starting structure obtained from previous equilibrium simulations.
Thermodynamic and dynamic properties were calculated over 700 000 steps
for the first set (and 900 000 steps for the
second set), previously equilibrated over 200 000 (or 300 000) steps.
The time step was 0.001 in reduced units and the time constant
of the Berendsen thermostat \cite{Be84} was 0.1 in reduced units.
The internal bonds between the particles in each dimer
remain fixed using the SHAKE \cite{Ryc77} algorithm, with a tolerance
of $10^{-12}$.

The stability of the system was checked by
analyzing the dependence of pressure on density and also by
visual analysis of the final structure, searching for cavitation.
The structure of the system was characterized using the intermolecular radial
distribution function, $g(r)$ (RDF), which does not take into account 
the correlation between atoms belonging to the same molecule. 
The diffusion coefficient was calculated using the slope of the least square
fit to the linear part of the mean square displacement, 
$<r^2(t)>$ (MSD), averaged over different time origins.

For analyzing the structure
we define the structural anomaly region as the region where the 
translational
order parameter $t$, given by
\begin{equation}
t\equiv\int_{0}^{\xi_{c}}|g(\xi)-1|d\xi,
\label{eq:trans}
\end{equation}
 \noindent decreases upon increasing density. Here 
$\xi\equiv r\rho^{1/3}$ is the  distance
$r$ in units of the
mean interparticle separation
$\rho^{-1/3}$, $\xi_c$ is the cutoff distance set to half of the
simulation box times $\rho^{-1/3}$, as in
Ref. \cite{Ol06b},
$g(\xi)$ is the radial distribution function as a function of the (reduced)
distance $\xi$ from a reference particle.
For an ideal gas  $g=1$
and $t=0$. In the crystal phase  $g\ne1$ over long distances and
$t$ is large.

\section{Results} \label{results}

\subsection{The phase-diagram}

\begin{figure}
\includegraphics[clip,scale=0.23]{schematic.eps}\includegraphics[clip,scale=0.23]{grcompress.eps}\includegraphics[clip,scale=0.23]{r2.eps}

\includegraphics[scale=0.15]{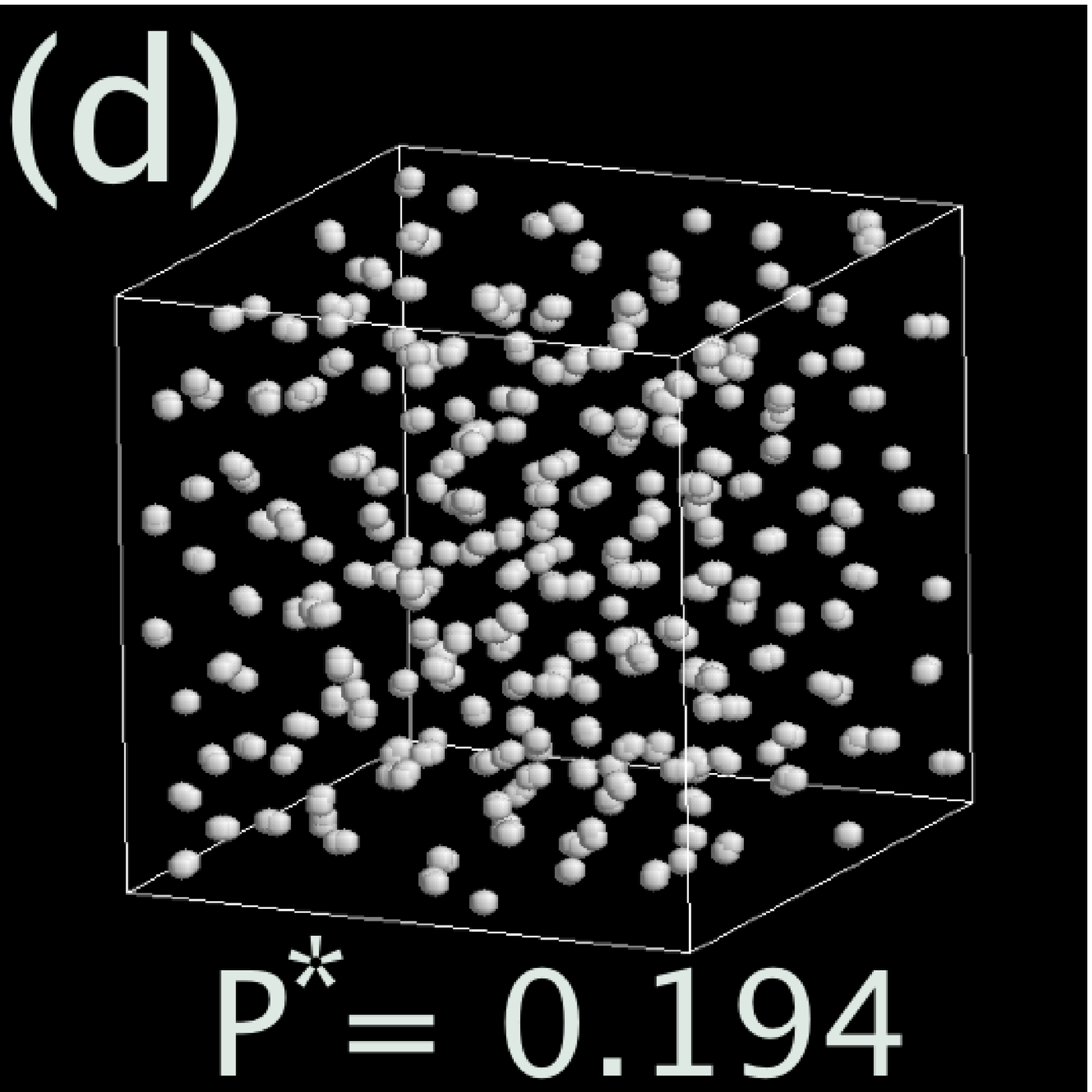}\includegraphics[scale=0.15]{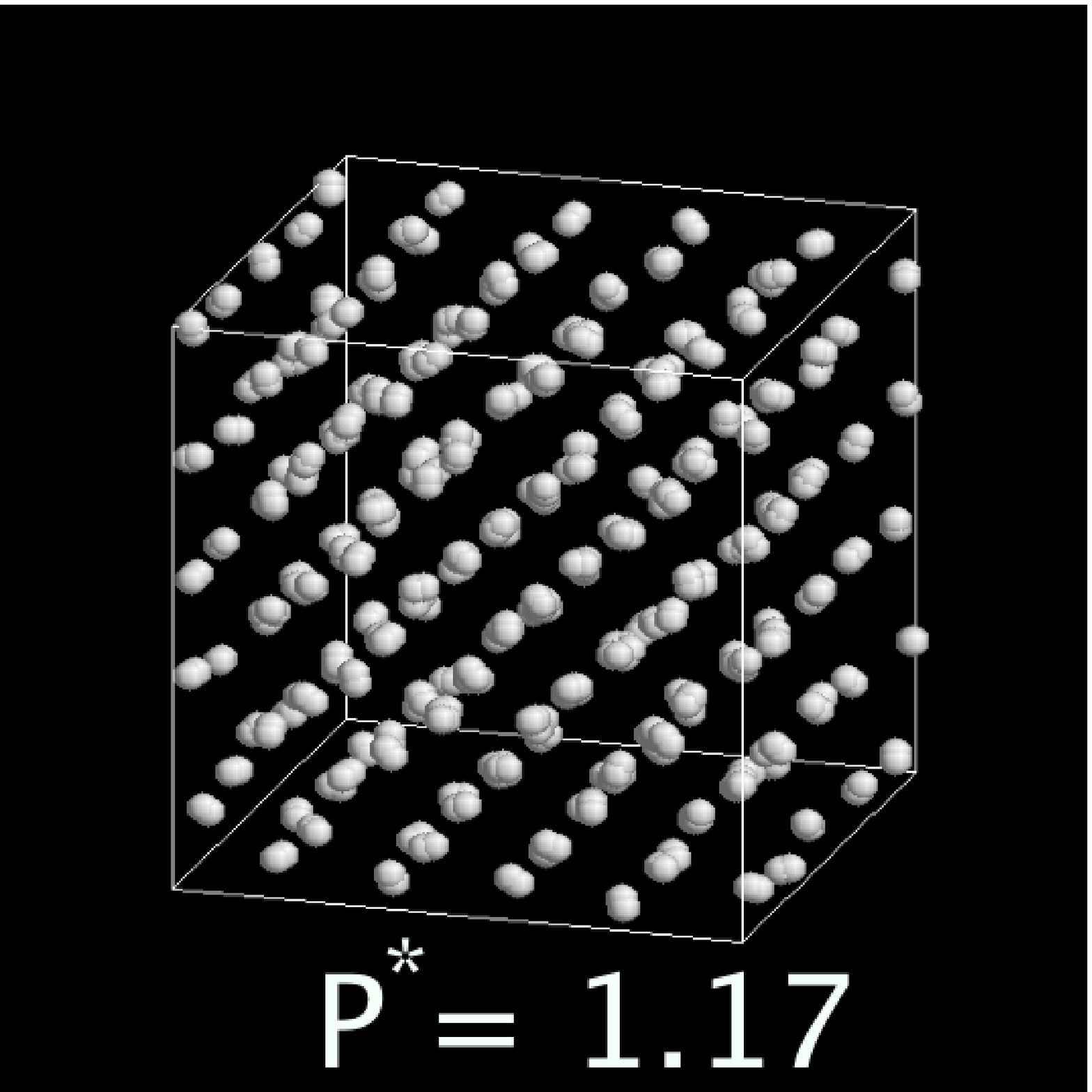}\includegraphics[scale=0.15]{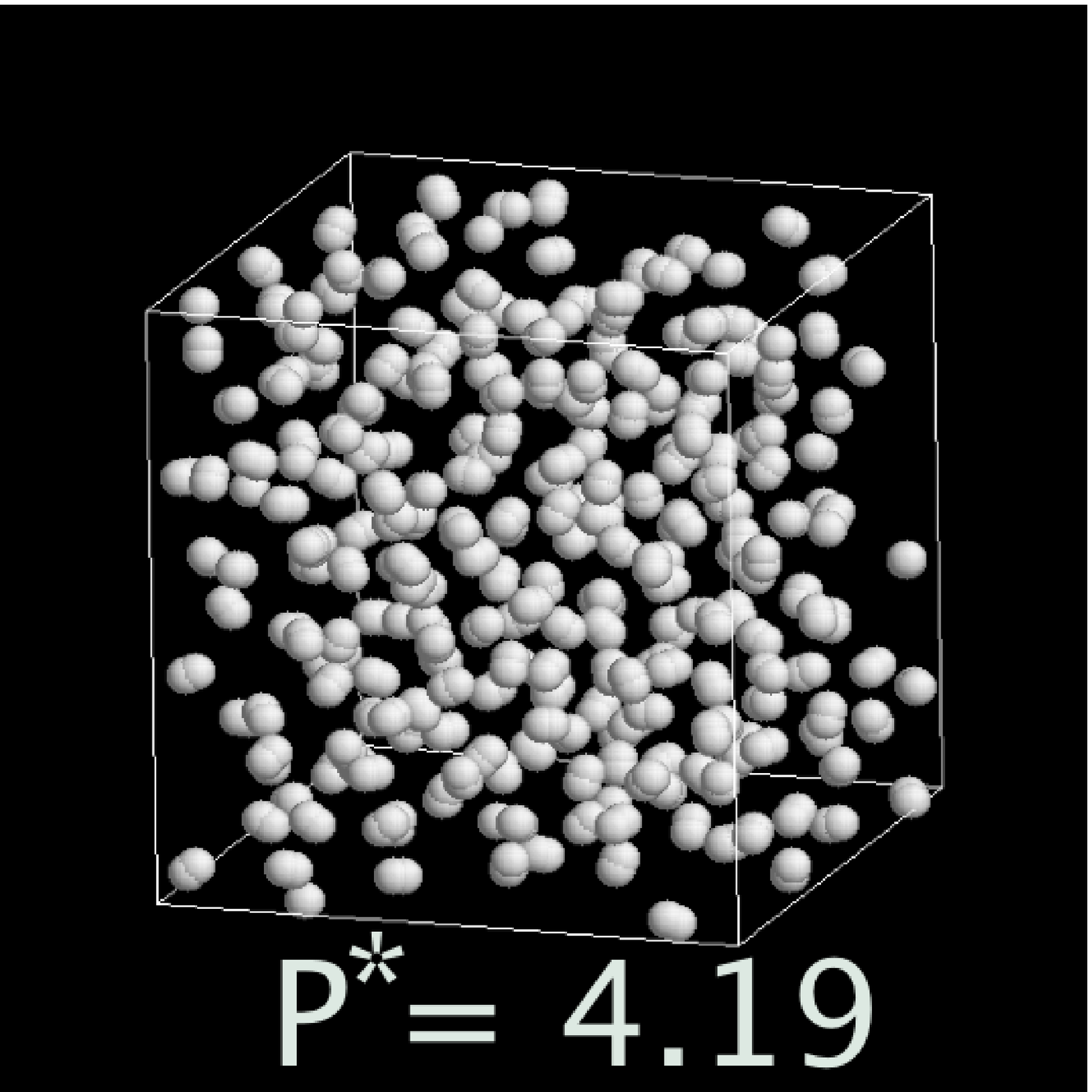}\includegraphics[scale=0.15]{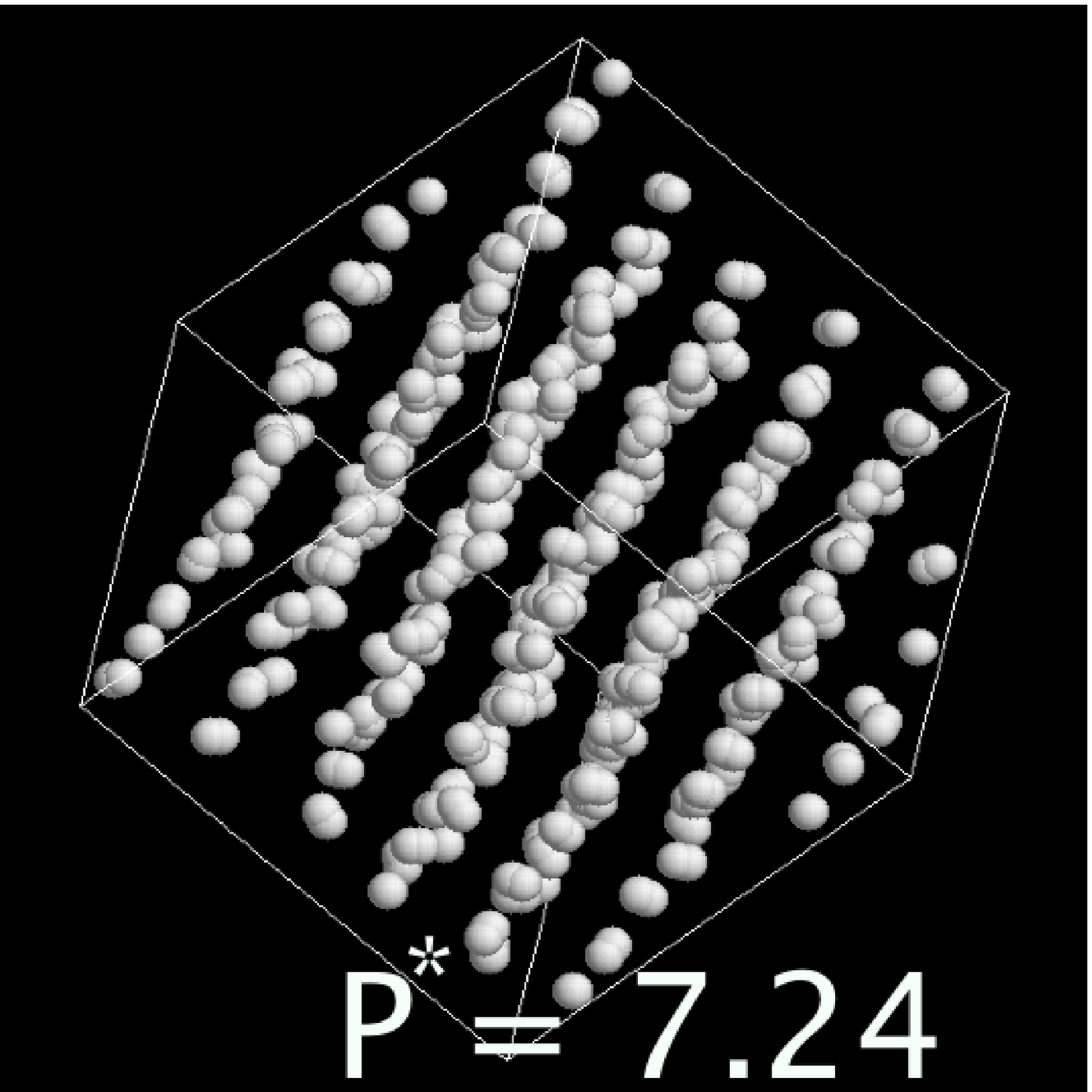}

\caption{(a) Reduced pressure  versus reduced
temperature phase-diagram showing the liquid and solid phases. The 
arrow at $T^*=0.4$ crosses the regions with $P^*=$ 0.194,
1.17, 4.19, and 7.24 illustrated in the radial distribution function.
(b) Radial distribution function versus reduced distance for
$T^{*}=0.4$ and $P^{*}=$ 0.194, 1.17, 4.19, and 7.24.
(c) Mean square displacement versus time for the four regions
illustrated in (a): fluid, solid, fluid,
and liquid crystal. With the slope of the MSD versus time
for  $P^{*}=$0.194, 1.17, 4.19, 7.24 we calculate the
diffusion coefficient that is zero for $P^*=1.17$.
(d) Snapshots of the configurations for $T^*=0.4$ and the same pressures
as in panel (a). 
\label{cap:compression}}
\end{figure}

Fig. \ref{cap:compression} shows (a) the pressure-temperature phase-diagram,
 (b) the radial distribution functions, 
(c) the mean square displacements and, finally, (d) snapshots of the system
at some relevant thermodynamic state points.  
The pressure-temperature phase diagram, illustrated in 
Fig. \ref{cap:compression}(a), 
displays at low temperatures
a low density solid phase, a high density solid phase, a 
low density fluid
phase, and a high density fluid phase. Near the boundaries of the high density
solid phase, a liquid crystal-like (LCL) phase was also identified, 
as discussed below.  

At intermediate temperatures
as the pressure is isothermally increased [following the
arrow in the figure \ref{cap:compression}(a)] the system
goes as follows: for low pressures the system is in the fluid phase, then as 
pressure increases it becomes solid, for higher pressures it becomes 
fluid again and
then liquid crystal for  even  higher pressures.  For very 
high pressures the system becomes solid and for even more
higher pressures it becomes fluid again (not shown in the 
figure). 

The nature of the phases was investigated from three different
ways: the radial distribution function, Fig. \ref{cap:compression}(b),
the mean square displacement,  Fig. \ref{cap:compression}(c), 
and the structural snapshots, Fig. \ref{cap:compression}(d).
While for $P^*=1.17$ and $7.24$ Fig. \ref{cap:compression}(d)
 shows ordered structures, for $P^*=0.194$ and  $4.19$ the 
structure is disordered, what could be typical for a liquid or a glass,
depending on the mobility of particles in the system.
According to the mean square displacement illustrated
in  Fig. \ref{cap:compression}(c), for $P^*=0.194$ and $P^*=4.19$ 
particles show high mobility, whereas
for $P^*=1.17$ particles do not move. This would 
be already expected from the corresponding radial distribution
function shown in Fig. \ref{cap:compression}(b).  However, for
$P^*=7.24$ particles move almost as fast as in
a liquid phase, but the corresponding radial distribution
function and structural snapshots are rather solid-like.

Which type of phase could be present in $P^*=7.24$? In order
to answer this question a movie with the evolution of the configurations 
was made.
This movie (available as supplementary material) 
shows that in this case particles actually move,
but only in a string-like fashion, what characterizes a
liquid crystal-like (LCL) phase.
Besides the presence of this liquid crystal phase, the 
dumbbell system also exhibit a fluid phase at
very high pressures. 

\subsection{The anomalies}

Similarly to the monomeric system 
previously studied \cite{Ol06a} anomalies were also 
found in its dumbbells version, considered in this work. We focused
in the three anomalies already found in the monomers case, i.e., the
density, diffusion, and structural anomalies, as described below.

(i)  The density anomaly is  
the unusual expansion of the system upon cooling
at constant pressure, a well-known effect which happens 
in water as discussed in the 
Introduction. In NPT-constant ensemble this anomaly is characterized by 
a maximum of the density along isobars in the density-temperature plane. 
Through thermodynamic relations \cite{Ku06} we are able to equivalently
detect this anomaly by searching for a minimum of the 
pressure along isochores in the pressure-temperature
phase-diagram. This was the technique used in this work since it is 
more suitable for the NVT ensemble.

\begin{figure}
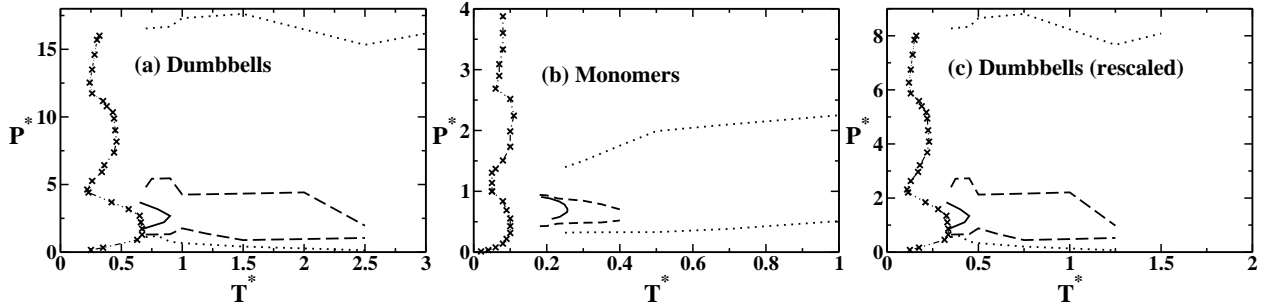

\includegraphics[clip=true,scale=0.25]{PT-DUMB.eps}\includegraphics[clip=true,scale=0.25]{PT-MONO-3.eps}\includegraphics[clip=true,scale=0.25]{PT-DUMB-2.eps}
\caption{Pressure versus temperature phase diagram for (a) dimeric and
(b) monomeric particles  interacting through the
potential illustrated in Fig. \ref{cap:shdumb}, and (c)
the rescaled dimeric results, in which pressure
and temperature from (a) are divided by 2 (see the text
for more details). The
results shown in panel (b) were adapted from
Refs. \cite{Ol06a} and \cite{Ol06b}.
The reentrant line with crosses in (a) represent the boundary between the fluid and 
the solid phases. The bold solid, dashed and dotted lines
represent the TMD, D extrema, and 
 the region of the structural anomaly
respectively, in both (a) and (b) panels.}
\label{cap:pressXT}
\end{figure}

The density anomaly region for both (a) dimeric and (b) monomeric systems
are shown in Fig. \ref{cap:pressXT} as solid bold lines. 
Both  dimeric and monomeric systems 
have the usual nose-shaped TMD line format,
also found in molecular models for water \cite{Er01,Ne01}
as well as for other isotropic potentials \cite{Xu06,Ya05,Ya06,Ol08b}.

For the dumbbells model the density anomaly region is located at
higher pressures and temperatures and occupies a much 
larger region in the pressure-temperature
phase-diagram than that observed for the the monomeric system. 
This can be evidenced by values for the maximum and minimum 
pressures and temperatures which enclose
the TMD line for both systems, as shown in Table \ref{tab:unica}.

(ii) We also studied the molecular mobility calculating the
diffusion coefficient using the the
mean-square displacement averaged over different initial times,
\begin{equation}
\langle \Delta r(t)^{2} \rangle = \langle [r(t_0+t)-r(t_0)]^2\rangle\; .
\end{equation}
Then the diffusion coefficient is obtained from the relation
\begin{equation}
D=\lim_{t\to\infty}\langle \Delta r(t)^{2} \rangle/6t \; .
\end{equation}

\begin{figure}
\includegraphics[clip=true,scale=0.3]{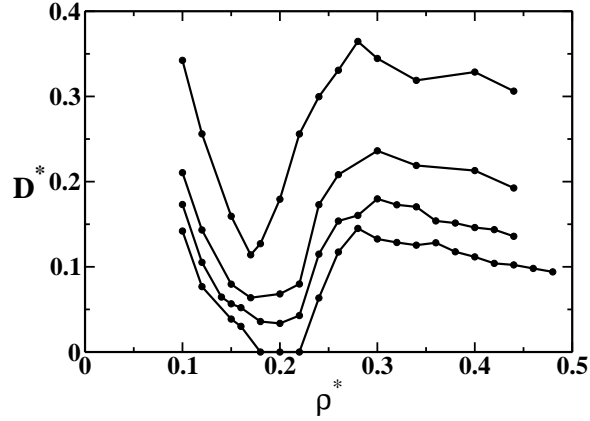}
\caption{Diffusion coefficient versus density at fixed temperatures, 
which are $T^{*}=$0.65, 0.75, 0.90, and 
1.20 from bottom to top.}
\label{cap:diff}
\end{figure}

Diffusion anomaly was also found in the dumbbells system. The  usual procedure
for detecting the dynamic anomaly region is to plot the diffusion 
coefficient versus
density for fixed temperatures \cite{Ol06a,Ol06b,Ku06,Xu06,Ol08b}
 as depicted in Fig. \ref{cap:diff}. From this figure
we see that for a certain range of densities the diffusion coefficient 
anomalously increases under increasing density, 
changing its slope from negative to positive
into this range. This determines a local maxima and minima for the 
$D(\rho)$ plot.
These extrema points can be mapped into a pressure-temperature
plane, determining the lines of extrema in the diffusion coefficient 
inside which
particles  move faster under 
compression -- or, equivalently, under increasing density.

Fig.~\ref{cap:pressXT}(a) illustrates the local diffusivity maxima and minima
as dashed lines. Comparison between Fig.~\ref{cap:pressXT}(a) and
Fig.~\ref{cap:pressXT}(b) indicates that the diffusion anomaly region
for dumbbells occupies a larger region in the p-T plane than
the diffusion anomaly region for the monomeric system. 
This can be evidenced by values for the maximum and minimum 
pressures and temperatures which enclose
the  diffusion anomalous region in  Table \ref{tab:unica}.

\begin{figure}
\includegraphics[clip=true,scale=0.3]{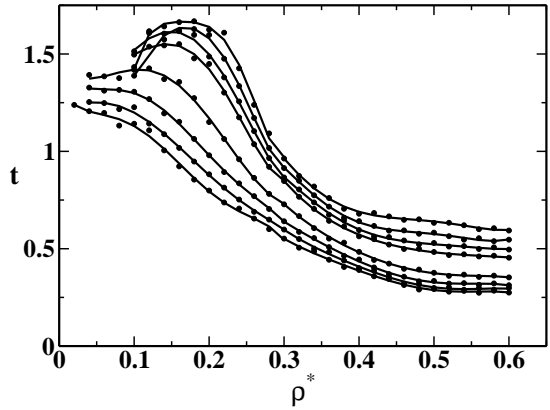}
\caption{Translational order parameter for the dumbbells system
against density for fixed temperatures,
which are $T^{*}=$ 0.7, 0.8, 0.9, 1.0, 1.5, 2.0, 2.5, and 3.0 
from top to bottom.}
\label{cap:trans}
\end{figure}

(iii) We also study the behavior of the translational order parameter.
We see from figure \ref{cap:trans} that $t$ decreases for increasing density
in a certain range of densities, contrary of what expected for a normal fluid.
The local maximum/minimum in the $t(\rho)$ plot
can be mapped into a pressure-temperature plane
(as discussed above) giving the bounding lines
for the region of structural anomaly in the $P$-$T$
phase diagram, i.e., dotted lines in Fig. \ref{cap:pressXT}.

Comparison between Fig.~\ref{cap:pressXT}(a) and
Fig.~\ref{cap:pressXT}(b) shows that 
the structural anomaly region is  much broader
for the dumbbells than for the monomers. It shall be stressed that
the dumbbells system can be found in the liquid phase at pressures
at high as $P^*=17.6$, which bounds the upper limit of the structural
anomaly region. On the other hand, the lower limit for 
the structural anomaly region of the dumbbells model
achieves pressures as low as 0.14
in reduced units, even lower than for the monomers case (which is 0.32 in 
reduced units).  

\begin{table}[htpb]
   \centering
   \topcaption{Maximum and minimum pressures and temperatures which bound (A)
the TMD line (solid in Fig. \ref{cap:pressXT}) in both systems, 
(B) dynamic anomaly region (bounded by dashed lines in 
Fig. \ref{cap:pressXT}) for
   both systems and (C) the structural anomaly region
    (bounded by dotted lines in Fig. \ref{cap:pressXT}) for
   both systems.} 
   \begin{tabular}{@{}  l  cc  @{} @{} ccc  @{} @{} ccc  @{}} 
      \\ \hline \hline
        & A & & & B & & & C & \\
        & Dimers & Monomers & & Dimers & Monomers & & Dimers & Monomers \\  \hline \hline
      $P^{*}$ max.      & 3.66 & 0.90 & &  5.42 & 0.94  & & $17.6$ &0.94   \\ \hline
      $P^{*}$ min.       &1.73 & 0.55 & &  0.90  & 0.43 & &$0.14$  & 0.43  \\ \hline
      $T^{*}$ max.       &0.90 & 0.26 & &  2.50  & 0.40 & &$>3.0$  & 1.00  \\ \hline
      $T^{*}$ min.      &0.65 & 0.18 & &  0.70  & 0.18 & &$<0.70$   & 0.25   \\ \hline \hline
      \bottomrule
\end{tabular}
\label{tab:unica}
\end{table}

In principle the differences between the monomeric and
the dimeric system could be interpreted by assuming that
a dimeric molecule is a monomeric with twice 
the potential $\epsilon$. In order to test if this would be the 
case, we rescaled pressure and temperature, $T^{**}=T^*/2$ and  $P^{**}=P^*/2$, and 
replot the diagram illustrated in Fig. \ref{cap:pressXT} (a)
obtaining Fig. \ref{cap:pressXT} (c). Even
in this case the dumbbell system shows remarkable differences 
when compared with the monomeric case. 

The addition of anisotropy changes the phase diagram of the fluid phase
being the differences significantly larger regarding the 
regions of the diffusion and structural anomalies. A proper choice
of parameters, maybe a fine tuning of the distance $\lambda$ between
the two particles of the same dimer, could lead to a specific
range in the pressure-temperature phase space of the anomalies.

\section{conclusion} \label{conclusion}

In resume, we have addressed the question if 
just including anisotropy in a core-softened potential
would lead to a modification in its anomalies and 
phases.  For that purpose we have 
investigated a dimeric version  of a previously studied monomeric model
whose particles were subject to a core-softened potential interaction 
\cite{Ol06a,Ol06b}. 

We have found a much richer phase-diagram for the dimers than that one 
obtained with the monomers. While for the monomeric system the solid
phase is at temperatures much lower than the temperatures of
the TMD region, for the dumbbell case the solid phase occupies
a much wider region in temperatures and terminates at
the edge of the TMD region. This indicates that 
the anisotropy favors certain ordering of the system at low
temperatures and low or intermediate pressures.
In the case of 
the dimeric particles two regions of solid phases separated by a liquid
crystal region were observed, suggesting that the 
two length scales together with the anisotropic arrangement
result in two solid phase densities.
Even more surprising a very high pressure a liquid phase 
is also observed. 

The thermodynamic, dynamic, and structural anomalous regions 
maintain the same hierarchy in the dimeric case as observed
in the monomeric system. The structural anomalous region occupies
a wider region in the pressure temperature phase-diagram when
compared with the diffusion anomalous region and this region 
occupies a wider region when compared with the density anomalous region.
This hierarchy is also observed in water \cite{Ne01,Ne02a}.
The dumbbell system, however, has pressures and temperatures
in these anomalous regions much larger than the ones observed
in the monomeric case. This behavior could be in principle explained
mapping the dumbbell in a monomer with twice the potential interaction
keeping the same interparticle distance. Since pressure and temperature
scale with the interparticle potential, increasing the potential
would lead to an increase in pressure and temperature.

In resume, the addition of an anisotropy in a core-softened potential
lead to a richer phase-diagram, but does not eliminates
the thermodynamic, dynamic, and structural anomalies present
in the isotropic system, but enlarges the solid
state region. This result indicates that the use
of dimers, trimers of even polymers of particles interacting with
this isotropic potentials is a promissing way to design
complex molecules which might lead to systems with
the same anomalies present in water or even other kinds of anomalous behavior.

\bibliography{Biblioteca}

\end{document}